\documentclass[preprint,aps]{revtex4}     % no pacs, no sectional numbering of eqns
\usepackage{graphicx}
\usepackage{amsmath,dsfont}
\def\beq{\begin{equation}}
\def\eeq{\end{equation}}
\def\beqnn{\begin{displaymath}}
\def\eeqnn{\end{displaymath}}
\def\bea{\begin{eqnarray}}
\def\eea{\end{eqnarray}}

\def\eno#1{Eq.~(\ref{#1})}

\def\fno#1{Fig.~\ref{#1}}

\def\tst{\textstyle}

\def\noi{\noindent}
\def\gam{\gamma}

\def\tta{\theta}

\def\apx{\approx}

\def\ptl{\partial}

\def\part#1#2{\frac{\ptl#1}{\ptl#2}}

\def\dint{\int\!\!\int}

\def\Tr{{\rm Tr\,}}

\def\gtwid{\mathrel{\raise.3ex\hbox{$>$\kern-.78em\lower1ex\hbox{$\sim$}}}}
\def\ltwid{\mathrel{\raise.3ex\hbox{$<$\kern-.78em\lower1ex\hbox{$\sim$}}}}

\def\hf{\frac{1}{2}}

\def\tshf{{\tst\hf}}

\def\ylm{Y_{\ell m}}
\def\Ylm{{{\cal Y}}_{\ell m}}

\def\tily{{\tilde Y}}

\def\typ0{\tily_{\ell+1,m}}

\def\tym0{\tily_{\ell-1,m}}

\def\Plp1{P_{\ell + 1}}
\def\Plm1{P_{\ell - 1}}

\def\llp1{\ell(\ell + 1)}
\def\tlp1{2\ell + 1}

\def\bJ{{\bf J}}

\def\ham{{\cal H}}

\def\ity{{\it y\/}\ }

\def\nhat{{\bf{\hat n}}}

\def\uhat{{\bf{\hat u}}}

\def\yhat{{\bf{\hat y}}}

\def\avg#1{\langle#1\rangle}
\def\bavg#1{\bigl\langle#1\bigr\rangle}

\def\ket#1{|#1\rangle}
\def\bra#1{\langle#1|}

\def\olap#1#2{\langle#1|#2\rangle}
\def\kb#1#2{\ket{#1}\bra{#2}}
\def\mel#1#2#3{\langle#1|#2|#3\rangle}

\begin{document}

\title{Wigner functions in the Higher-Spin Einstein-Podolsky-Rosen-Bohm Experiment}
\author{Anupam Garg}
\email[e-mail address: ]{agarg@northwestern.edu}
%\email{agarg``at"northwestern.edu}
\affiliation{Department of Physics and Astronomy, Northwestern University,
Evanston, Illinois 60208}

\date{\today}

\begin{abstract}
The spin-$j$ extension of Bohm's version of the Einstein-Podolsky-Rosen experiment is
is analysed in terms of the Wigner function when the two spins are in a singlet state.
This function is calculated for all $j$, and
it is shown that just as Bell inequalities are violated with undiminished range and
magnitude for arbitarirly large $j$, this function does not become less negative. On the contrary,
the oscillations between positive and negative values grow both in frequency and amplitude.
It is argued that this is an alternative way to grasp the approach to classical behavior
with increasing quantum number.

\vskip20pt\noi
Keywords: Einstein-Podolsky-Rosen experiment, Bell inequalities, Wigner function, spin Wigner function.
\end{abstract}

%\pacs{75.50.Xx, 75.60.Jk, 76.20.+q, 75.45.+j}
\maketitle

%\section{Introduction}
%\label{intro}

Bell's analysis of Bohm's spin-1/2 version of the Einstein-Podolsky-Rosen (EPR) experiment\,%
\cite{epr,bohm,bell}
and its photonic analog have been the subject of so much work and discussion that any attempt to
say something new is fraught with irrelevance. The subject has passed from
physics to popular culture, and my recent Google search (approximately 16:00 UTC, Oct.~28, 2019) on `Bell basis'
produced more hits than `Elvis Presley' and `Mahatma Gandhi' ($178 \times 10^6$ vs.~$108 \times 10^6$
and $54 \times 10^6$). Within physics, from its origin as a gedanken experiment and an academic
metaphysical puzzle, it has turned into a central theme in quantum information science and
communication\,%
\cite{bb84,eke91,nc02}.
Yet, despite these applications, the conceptual problems of objective local realism, counterfactual
definiteness, and the system-observer split have lost none of their capacity to befuddle and we are
no closer today to reconciling the ``spooky actions at a distance" with our everyday classical
experience and world view than we ever were.
The spin-1/2 system is of course profoundly nonclassical, and so in the 1980's and 90's several
investigations were undertaken of the spin-$j$ version of this experiment in the belief that they
would reveal a gradual recovery of classical physics with increasing $j$. However, Garg and Mermin\,%
\cite{gm82,gm83} found Bell inequalities that were violated with a range that did not diminish at all
as $j \to \infty$, although the magnitude of violation did, and did so exponentially. Gisin and Peres\,%
\cite{gp92} then found an inequality for which the magnitude of violation did not vanish either.
This inequality is based on an algebraic construction of a dichotomic operator which does not correspond
to more accessible measurables such as spin components. These papers show that large quantum numbers do
not necessarily behave more classically, although the nonclassical features are seen only via specialized
and arguably contrived constructions.

In this paper, we approach this question by finding a simple closed form for the exact Wigner function of
the spin-$j$ singlet state. The Wigner function for {\it pq\/} systems, i.e., for degrees of
freedom that can be cast as canonically conjugate position and momentum variables, is even older than
the EPR paper\,%
\cite{epw32}
and has proven to be a very fruitful concept in understanding nonclassical aspects of quantum systems,
and has given rise to practical methods and approximation schemes for dynamical evolution\,%
\cite{wk18}.
For spin systems, the use of the Wigner function has so far been largely conceptual\,%
\cite{strato, kutzner,agarwal,sar78,vgb89,lbg13}, 
and even the applications have been somewhat formal\,
\cite{bgls}.
Our work may well strike some readers as having a similar character, but we hope that it will lead
to new ways of thinking about and working with spin.

We find that the Wigner function of the spin-$j$ singlet does not become less negative with increasing
$j$ in any sense. Instead the oscillations between positive and negative values become ever more rapid and
wilder as $j$ increases.
(See \eno{rho_W_yy2} and \fno{wig_fns_epr.eps} below.)
This is in accord with the Garg-Mermin and Gisin-Peres conclusion that increasing
quantum number does not equate to increasing classicality, but its statement in terms of the Wigner
function is more direct and gives us an alternative way of thinking about it.

For a {\it pq\/} system in one dimension in a state with the density matrix $\rho$, the expectation value of
any operator $F$ is given by
\beq
\avg{F}_{\rho}
  = \Tr(\rho F)
  = \dint \frac{dp\,dq}{2\pi}\, W_{\rho}(p,q) \Phi^W_F(p,q),
  \label{avg_F_pq}
\eeq
where $W_{\rho}$ is the Wigner function, and $\Phi^W_F$ is the Weyl transform of $F$\,%
\cite{weyl,moy}.
In fact, the Wigner function is itself the Weyl transform of the density operator, and \eno{avg_F_pq} is
a special case of the {\it traciality theorem\/}
\beq
\Tr(FG)
  = \dint \frac{dp\,dq}{2\pi}\, \Phi^W_F(p,q) \Phi^W_G(p,q)
  \label{avg_FG_pq}
\eeq
for any two operators $F$ and $G$. All the Weyl transforms are functions on phase space, and were the Wigner
function not negative, quantum mechanics could be cast as a classical stochastic theory.

The first question that must be confronted for spin is what the relevant phase space is. For a single
spin, we argue that the natural choice is the unit sphere, $S^2$, points on which are directions $\nhat$ in
three-dimensional space\,%
\cite{woo87}.
Just as for a {\it pq\/} system the variables $p$ and $q$ are the classical
counterparts of the operators $p^{\rm op}$ and $q^{\rm op}$, for spin,
the components of $\nhat$ correspond to the components of $\bJ/j$, where $\bJ$ is the spin operator,
and just as for the former the Wigner function aims to give the joint distribution for the noncommuting
$p^{\rm op}$ and $q^{\rm op}$, for the latter it aims to give the joint distribution of the noncommuting
operators $J_{\mu}/j$, ($\mu = x, y, z$). All spin-$j$ operators are matrices of order $2j+1$, and so by the
Cayley-Hamilton theorem, they can be
written as polynomials of degree $2j$ or less in the $J_{\mu}$. This is also true of any Hamiltonian
$\ham(\bJ)$, whence it follows that $\bJ\cdot\bJ$ is a constant of motion, allowing us to think of the classical spin
as living and moving on $S^2$.

Unlike {\it pq\/} systems for which there is a direct definition of the Weyl transform, for spin an
indirect approach must be adopted. We demand that the Weyl map $\Phi^W_F(\nhat)$ of any operator $F(\bJ)$ be linear,
take the unit operator into the unit function, hermitean operators into real-valued functions,
and be covariant under rotations. Most importantly it should obey the {\it traciality condition\/}\,%
\cite{vgb89,lbg13}
\beq
\avg{FG}_{\rm qm} = \bavg{\Phi^W_F(\nhat) \Phi^W_G(\nhat)}_{\nhat},
  \label{avg_FG_spin}
\eeq
where $\avg{F}_{\rm qm} = (2j+1)^{-1}\Tr(F)$ and $\avg{f(\nhat)}_{\nhat} = \int f(\nhat)\, d^2\nhat/4\pi$.
This fixes the map completely.
It then follows from the requirement that $\avg{F}_{\rho} = \Tr(\rho F)$ that the Wigner function $W_{\rho}(\nhat)$
is a normalizing factor $(2j+1)/4\pi$ times the Weyl transform $\Phi^W_{\rho}(\nhat)$\,%
\cite{no_marginal}.

A prescriptive algorithm for finding $\Phi^W_F$ is as follows \cite{lbg13}. We first find the Q transform\,%
\cite{hus40,sud63,gla63}
\beq
\Phi^Q_F(\nhat) = \mel{\nhat}{F}{\nhat},
  \label{Q_fn}
\eeq
where $\ket{\nhat}$ is a spin coherent state\,%
\cite{radcliffe,arecchi},
i.e., the eigenstate state $\ket{j,j}_{\nhat}$ of $\bJ\cdot\nhat$ with eigenvalue $j$, and thus the state
with maximum spin projection along the direction $\nhat$. Then,
\beq
\Phi^W_F(\nhat) = \int d^2\nhat' \, M^{WQ}(\nhat,\nhat') \Phi^Q_F(\nhat'),
  \label{Q_to_W_map}
\eeq
where,
\beq
M^{WQ}(\nhat,\nhat')
  = \sum_{\ell = 0}^{2j} \sum_{m = - \ell}^{\ell}
       S^{-1}_{j\ell}
         \ylm(\nhat) \ylm^*(\nhat'),
  \label{Q_to_W_kernel}
\eeq
with
\beq
S_{j\ell} = \prod_{k=0}^{\ell} \Bigl( \frac{2j + 1 -k}{2j + 1 + k} \Bigr)^{1/2}.
  \label{S_jell}
\eeq
The origin of the kernel $M^{WQ}$ is that the Weyl and Q transforms of the spherical harmonic tensor
operator $\Ylm(\bJ)$ are both proportional to $\ylm(\nhat)$ by rotational covariance, and the ratio
of the proportionality constants is $S^{-1}_{j\ell}$. The inverse kernel $M^{QW}$ that maps
$\Phi^W_F$ into $\Phi^Q_F$ is obtained by replacing $S^{-1}_{j\ell}$ with $S_{j\ell}$ in \eno{Q_to_W_kernel}.

We now apply this algorithm to the singlet state, $\ket{\phi}$. Readers who do not wish to see how this is
done should skip to the result, \eno{rho_W_yy2}. It should be kept in mind that we now have a two-spin system, so the
phase space is $S^2 \times S^2$, points in which are specified by a pair of unit vectors $(\nhat_1, \nhat_2)$,
the subscripts 1 and 2 labelling the two particles. We have
\beq
\ket{\phi} = \frac{1}{\sqrt{2j+1}} \sum_{m=-j}^j (-1)^{j-m} \ket{j,m}_{1,\uhat} \otimes \ket{j,-m}_{2,\uhat},
  \label{ans_singlet}
\eeq
where $\ket{j,m}_{k,\uhat}$ ($k=1,2$) is the eigenstate of $\bJ_k\cdot\bJ_k$ and $\bJ_k\cdot\uhat$ with eigenvalues
$j(j+1)$ and $m$. Because $\ket{\phi}$ is rotationally invariant, \eno{ans_singlet} is valid for any
quantization axis $\uhat$. The Q transform of the density operator $\rho = \kb{\phi}{\phi}$ is given by
\beq
\Phi^Q_{\rho}(\nhat_1, \nhat_2) =  |\olap{\phi}{\nhat_1, \nhat_2}|^2.
\eeq
Taking $\uhat = \nhat_1$ in \eno{ans_singlet}, we obtain
\beq
\Phi^Q_{\rho}(\nhat_1, \nhat_2)
  = \frac{1}{2j+1} |\olap{-\nhat_1}{\nhat_2}|^2
  = \frac{1}{2j+1} \bigl[\tshf (1 -\nhat_1\cdot \nhat_2)\bigr]^{2j},
  \label{Q_fn_phi}
\eeq
where the last equality follows from a well-known result for the Wigner rotation matrix element
${\cal D}^{j}_{jj}(\yhat,\tta)$ for a rotation about $\yhat$ through an angle $\tta$\,%
\cite{ll'_rot_mat}

The next step is to expand $\Phi^Q_{\rho}$ in Legendre polynomials of $\nhat_1\cdot\nhat_2$. The
required analysis is straightforward, and we
get
\beq
\Phi^Q_{\rho}(\nhat_1,\nhat_2)
  = \frac{1}{2j+1}
     \sum_{\ell = 0}^{2j} (2\ell + 1) A_{j\ell} P_{\ell}(-\nhat_1\cdot\nhat_2),
   \label{Q_singlet}
\eeq
where
\beq
A_{j\ell} = \frac{ \bigl[(2j)!\bigr]^2}{(2j-\ell)!\, (2j + \ell +1)!}.
  \label{ajl}
\eeq
Using the addition theorem for spherical harmonics, we thus obtain
\beq
\Phi^Q_{\rho}(\nhat_1,\nhat_2)
  = \frac{4\pi}{2j+1}
     \sum_{\ell = 0}^{2j}\sum_{m=-\ell}^{\ell}
          A_{j\ell}
              \ylm(\nhat_1) \ylm^*(-\nhat_2).
  \label{rho_Q_yy}
\eeq

The last step is to apply the map (\ref{Q_to_W_map}) to $\Phi^Q_{\rho}$. This must be done
for both $\nhat_1$ and $\nhat_2$. Invoking the orthonormality of the $\ylm$'s, we find that
\beq
\Phi^W_{\rho}(\nhat_1,\nhat_2)
  = \frac{4\pi}{2j+1}
     \sum_{\ell = 0}^{2j}\sum_{m=-\ell}^{\ell}
          S^{-2}_{j\ell} A_{j\ell}
              \ylm(\nhat_1) \ylm^*(-\nhat_2).
  \label{rho_W_sum}
\eeq
But, as is easily checked,
\beq
S^{-2}_{j\ell} A_{j\ell} = \frac{1}{2j+1}.
\eeq
Hence,
\beq
\Phi^W_{\rho}(\nhat_1,\nhat_2)
  = \frac{4\pi}{(2j+1)^2}
     \sum_{\ell = 0}^{2j}\sum_{m=-\ell}^{\ell}
              \ylm(\nhat_1) \ylm^*(-\nhat_2).
  \label{rho_W_yy}
\eeq
Multiplying by the normalization factor $[(2j+1)/4\pi]^2$, we obtain the especially simple result
\beq
W_{\rho}(\nhat_1,\nhat_2)
  = \frac{1}{4\pi}
     \sum_{\ell = 0}^{2j}\sum_{m=-\ell}^{\ell}
              \ylm(\nhat_1) \ylm^*(-\nhat_2).
  \label{rho_W_yy2}
\eeq
By employing the addition theorem in reverse we can write it as
\beq
W_{\rho}(\nhat_1,\nhat_2)
  = \frac{1}{(4\pi)^2}
     \sum_{\ell = 0}^{2j} (2\ell +1) P_{\ell}(-\nhat_1\cdot\nhat_2),
              \label{rho_W_P_ell}
\eeq
which manifests the rotational invariance of the singlet state since it depends only on the angle between
$\nhat_1$ and $\nhat_2$.
%
%It is useful to define
%\beq
%x_{12} = -\nhat_1\cdot\nhat_2.
%\eeq

%
\begin{figure}[h]
\centering
\includegraphics[width=6in]{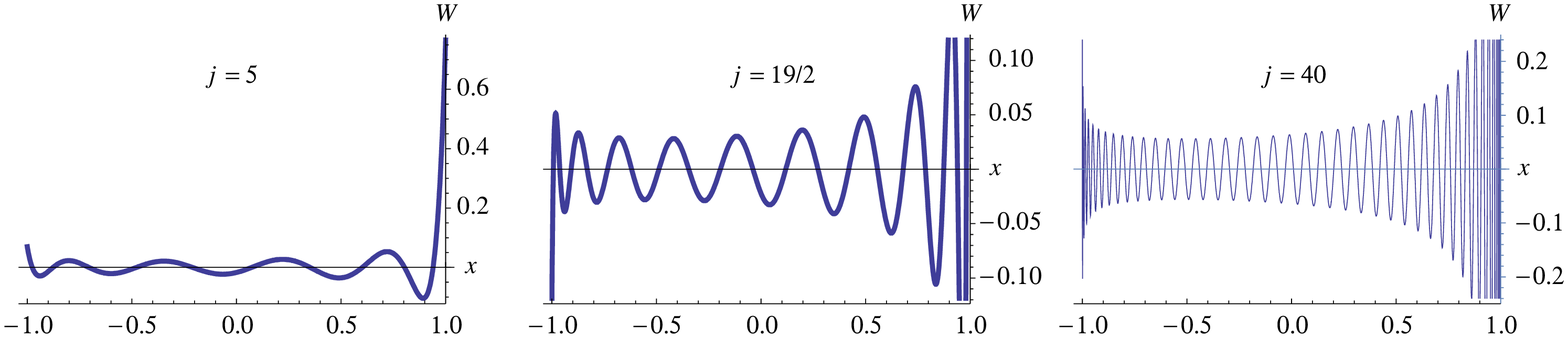}
\vskip0.1cm
\caption{Wigner functions for $j=5$, $19/2$, and $40$, as a function of $x = -\nhat_1\cdot\nhat_2$.
         The scales on the \ity axes should be noted. For $j=19/2$ and $40$, large portions of the vertical range
         are not shown.}
\label{wig_fns_epr.eps}
\end{figure}
Before analyzing \eno{rho_W_yy2} further, we plot $W_{\rho}$ for three different values of $j$
in \fno{wig_fns_epr.eps}.
As is evident, the function is not positive, and as asserted earlier, both the frequency and the
amplitude of the oscillations increase with $j$. We can show analytically that:

(A) $W_{\rho}(\nhat_1, -\nhat_1) = (2j+1)^2/(4\pi)^2$.

(B) $W_{\rho}(\nhat_1, \nhat_1) = (-1)^{2j} (2j+1)/(4\pi)^2$.

(C) The scale of the oscillations is $\sim j^{1/2}$.

(D) The first zero of $W_{\rho}$ next to its main peak is at a distance $\sim 7.34/(2j+1)^2$ from
$x = -\nhat_1\cdot\nhat_2 = 1$. Thus this peak has a width inversely proportional to the height,
consistent with the requirement that the total probability be unity.

Property (A) follows from direct evaluation of the sum in \eno{rho_W_P_ell} and the fact
$P_n(1) = 1$. For the others, we appeal to the Christoffel-Darboux theorem\,%
\cite{ab_steg,szego},
which gives this sum as a particular case. The result we need is\,%
\beq
S_j(x) \equiv \sum_{\ell = 0}^{2j} (2\ell +1) P_{\ell}(x)
  = \frac{2j+1}{1-x} [P_{2j}(x) - P_{2j+1}(x)].
\eeq
Since $P_n(-1) = (-1)^n$, $S_j(-1) = (-1)^{2j} (2j+1)$, which gives property (B).
For properties (C) and (D), we rely on standard asymptotic forms for $P_n(x)$ for large $n$.
With $\cos\gam = -\nhat_1\cdot\nhat_2$, we obtain
\beq
W_{\rho}(\cos\gam)
  \apx \frac{(2j+1)}{(4\pi)^2} \frac{1}{1 - \cos\gam}
     \Bigl(\frac{\gam^3}{\sin\gam} \Bigr)^{1/2}
        J_1((2j+1)\gam),
\eeq
where $J_1$ is the Bessel function of order 1, and its asymptotic behavior for large arguments gives us
(C). The first nontrivial zero of $J_1$ is at $j_{1,1} = 3.8317$. This gives the location of the first zero
of $W_{\rho}$ at
\beq
1 - x \apx \frac{j_{1,1}^2}{2(2j+1)^2}
   = \frac{7.34}{(2j+1)^2}.  \label{first_zero}
\eeq
This is property (D). We also find that the frequency of the oscillations grows with $j$ as $j^1$.


\begin{thebibliography}{99}
%
\bibitem{epr}
A.~Einstein, B.~Podolsky, and N.~Rosen,
Can the quantum-mechanical description of physical reality be considered complete?
Phys.\ Rev.\ {\bf 47}, 777 (1935).
%
\bibitem{bohm}
D. Bohm,
{\it Quantum Theory}
(Prentice-Hall, Englewood Cliffs, New Jersey, 1951), pp.~614--619.
%
\bibitem{bell}
Bell, J.~S.
On the Einstein-Podolsky-Rosen Paradox,
{\it Physics}\ {\bf 1}, 195--200 (1964).
%
\bibitem{bb84}
C.~Bennett and G.~Brassard,
{\it Quantum Cryptography: Public Key Distribution and Coin Tossing\/}
(in Proc.\ IEEE International Conference on Computers, Systems and Signal
Processing, Bangalore, Dec.~1984),
IEEE, New York (1984).
%
\bibitem{eke91}
A.~Ekert,
Quantum cryprography based on Bell's theorem,
Phys.\ Rev.\ Lett.\ {\bf 67}, 661--663 (1991).
%
\bibitem{nc02}
M.~Nielsen and I.~Chuang,
{\it Quantum Computation and Quantum Information\/},
Cambridge University Press, Cambridge, 2002.
%
\bibitem{gm82}
Garg, A. and Mermin, N.~D.,
Bell inequalities with a range of violation that does not diminish as the spin becomes
arbitrarily large,
{\it Phys.\ Rev.\ Lett.}\ {\bf 49}, 901--904 (1982).
%
\bibitem{gm83}
Garg, A. and Mermin, N.~D.,
Local Realism and Measured Correlations in the spin-$s$ Einstein-Podolsky-Rosen
Experiment,
{\it Phys.\ Rev.\ D}\ {\bf 27}, 339--348 (1982).
%
\bibitem{gp92}
N.~Gisin and A.~Peres,
Maximal violation of Bell's inequality for arbitrarily large spin,
Phys.\ Lett.\ A {\bf 162}, 15--17 (1992).
%
\bibitem{epw32}
E.~Wigner,
On the	Quantum Correction for Thermal Equilibrium,
Phys.\ Rev.\ {\bf 40}, 749--759 (1932).
%
\bibitem{wk18}
J.~Weinbub and D.~K. Ferry,
Recent Advances in Wigner Function Approaches,
Appl.\ Phys.\ Rev. {\bf 5}, 041104 (2018).
%
\bibitem{strato}
R.~L. Stratonovich, Sov. Phys. JETP {\bf 31}, 1012 (1956).
%
\bibitem{kutzner}
J.~Kutzner,
Z.\ Phys.\ {\bf 259}, 177 (1973).
%
\bibitem{agarwal}
G.~S. Agarwal, Phys. Rev. A {\bf 24}, 2889 (1981).
%
\bibitem{sar78}
B.~S. Shastry, G.~S. Agarwal, and I.~Rama Rao, Pramana {\bf 11}, 85 (1978).
%
\bibitem{vgb89}
J.~C. Varilly and J.~M. Gracia-Bondia, Annals of Phys. (NY) {\bf 190}, 101 (1989).
%
\bibitem{lbg13}
F.~Li, C.~Braun, and A.~Garg,
The Weyl-Wigner-Moyal formalism for spin,
Europhys. Lett. {\bf 102}, 60006 (2013).
%
\bibitem{bgls}
C.~Braun, F.~Li, M.~Stone, and A.~Garg,
J.\ Math.\ Phys.\ {\bf 56}, 122106 (2015).
%
\bibitem{weyl}
H.~Weyl, {\it The Theory of Groups and Quantum Mechanics\/}, Dover Publications, New York (1950) [translation
of {\it Gruppentheorie und Quantenmechanik\/}, Hirzel Verlag, Leipzig (1928)]. See Chap.~II, Sec.~11 and Chap.~IV, Sec.~14.
%
\bibitem{moy}
J.~E. Moyal, Proc. Cambridge Philos. Soc. {\bf 45}, 99 (1949).
%
\bibitem{woo87}
A rather different approach to the spin Wigner function is taken in
W.~Wooters,
A Wigner-function formulation of finite state quantum mechanics,
Ann.\ Phys.\ (NY) {\bf 176}, 1--21 (1987).
Wooters is more concerned with the discreteness of the spectrum of the $J_{\mu}$, and with
preserving the concept of a marginal. His phase space is a square $N \times N$ lattice of
points with $N = 2j+1$ if $N$ is a prime number. If it is not, the phase space has to be conceived of
as a product of spaces associated with the prime factors of $N$. Further, there is no notion of
covariance under rotations, and the connection between spin and rotations is secondary
and largely lost.
%
\bibitem{no_marginal}
For {\it pq\/} systems, the fact that $W_{\rho}(p,q)$ is not a respectable distribution shows up in its
nonpositivity. For spin systems, not only is this so, but the notion of a marginal makes no sense
either. Integrating over the other directions to obtain the probability distribution of $J_z/j$, say,
yields a function of a continuous variable, with no hint that the true spectrum of this
operator is discrete. $W_{\rho}(\nhat)$ is only good for calculating the averages of observables
via \eno{avg_FG_spin}.
%
\bibitem{hus40}
K.~Husimi,
Proc.\ Phys.\ Math.\ Soc.\ Jpn.\ {\bf 22}, 264--314 (1940).
%
\bibitem{sud63}
E.~C.~G.~Sudarshan,
Phys.\ Rev.\ Lett.\ {\bf 10}, 277 (1963).
%
\bibitem{gla63}
R.~J. Glauber,
Phys.\ Rev. {\bf 131}, 2766 (1963).
%
\bibitem{radcliffe}
J.~M. Radcliffe,
J.\ Phys.\ A: Gen.\ Phys.\ {\bf 4}, 313 (1971).
%
\bibitem{arecchi}
F.~T. Arecchi, E.~Courtens, R.~Gilmore, and H.~Thomas,
Phys.\ Rev.\ A {\bf 6}, 2211 (1972).
%
\bibitem{ll'_rot_mat}
L.~D. Landau and E.~M. Lifshitz,
{\it Quantum Mechanics (Nonrelativistic Theory)\/}, 3rd ed., Pergamon, Oxford, 1977;
see Eq.~(58.26).
%
\bibitem{ab_steg}
Abramowitz, M.~\& and Stegun, I.~A.
{\it Handbook of Mathematical Functions\/}, Nat'l Bureau of Standards Applied
Math.\ Series 55 (1964). See Eq.\,22.12.1.
%
\bibitem{szego}
G. Szego, {\it Orthogonal polynomials\/}, Amer.\ Math.\ Soc.\ Colloquium
Publications, {\bf 23}, (1939); Sec.~3.2.
%
%\bibitem{lg85}
%Leggett, A.~J.~\& Garg, A.
%Quantum mechanics versus macroscopic realism: is the flux there when nobody looks?
%{\it Phys.\ Rev.\ Lett.}\ {\bf 54}, 857--860 (1985).
%%
%\bibitem{ms82}
%Mermin, N.~D.~\& Schwarz, G.~M.
%Joint distributions and local realism in the higher-spin Einstein-Podolsky-Rosen
%experiment,
%{\it Found.\ Phys.}\ {\bf 12}, 101--135 (1982).
%
%\bibitem{jens_me}
%Very recently, Jens Koch and the author (to be published), have found an exact asymptotic
%form for this map in terms of exponentials of Faulhaber polynomials of the Laplace-Beltrami
%operator on the unit sphere.
%%
%\bibitem{W_osc}
%$(4\pi)^2 W_{\rho}$ equals $(2j+1)^2$ at $x=1$, $(-1)^j (2j+1)$ at $x=-1$, the scale of
%the oscillations is $\sim j^{1/2}$, and the first zero of $W_{\rho}$ is at a distance
%of $7.34/(2j+1)^2$ from the main peak at $x = 1$. These results, and asymptotic forms for
%$W_{\rho}$ will be published elsewhere.
%%
%\bibitem{radcliffe}
%J.~M. Radcliffe,
%J.\ Phys.\ A: Gen.\ Phys.\ {\bf 4}, 313 (1971).
%%
%\bibitem{arecchi}
%F.~T. Arecchi, E.~Courtens, R.~Gilmore, and H.~Thomas,
%Phys.\ Rev.\ A {\bf 6}, 2211 (1972).
%%
%\bibitem{ag83}
%A.~Garg, Phys. Rev. D {\bf 28}, 785 (1983).
%%
%\bibitem{ag_thesis}
%A.~Garg, Ph.D. thesis, Cornell University, 1983.
%%
\end{thebibliography}
\end{document}